\documentclass[%
 reprint,
 amsmath,amssymb,
 aps,
 pre,
]{revtex4-2}

\usepackage{graphicx}% Include figure files
\usepackage{dcolumn}% Align table columns on decimal point
\usepackage{bm}% bold math
\usepackage{subfigure}
\usepackage{mathtools}
\newcommand\numberthis{\addtocounter{equation}{1}\tag{\theequation}}

\begin{document}

%\preprint{APS/123-QED}

\title{Dynamics of a chiral swimmer sedimenting on a flat plate}% Force line breaks with \\

\author{Federico Fadda}
 %\email{fede.fadda1110@gmail.com}
\author{John Jairo Molina}%
\author{Ryoichi Yamamoto}
\email{ryoichi@cheme.kyoto-u.ac.jp}
\affiliation{Department of Chemical Engineering, Kyoto University, Kyoto 615-8510, Japan}%


%\affiliation{
%\affiliation{
 %Third institution, the second for Charlie Author
%\affiliation{%


\date{\today}% It is always \today, today,

\begin{abstract}
Three-dimensional simulations with fully resolved hydrodynamics are performed to study the dynamics of a single squirmer under gravity, in order  to clarify its motion in the vicinity of a flat plate. Different dynamics emerge for different gravity strengths. In a moderate gravity regime, neutral squirmers and pullers eventually stop moving and reorient in a direction perpendicular to the plate; pushers, instead, exhibit continuous motion in a tilted direction. In the strong gravity regime, all types of squirmers sediment and reorient perpendicularly to the plate. In this study, the chirality is introduced to model realistic micro-swimmers, and its crucial effects on the swimmer dynamics are presented.
%\begin{description}
%\end{description}
\end{abstract}

\maketitle

%\tableofcontents

\section{\label{sec:level1_1}INTRODUCTION}

The motion of swimming microorganisms, like algae and bacteria, is characterized by three basic features: (1) they swim at low Reynolds number, where viscous forces dominate inertial ones, (2) the net force and torque exerted on the squirmer are zero and (3) the presence of boundaries and confinements can significantly affect their swimming behaviour. In nature, microorganisms regularly encounter surfaces, like sperm swimming in the mammalian tract or bacteria forming biofilms on surfaces for spreading. Many studies related to this problem have been published \cite{lauga1,marchetti1,gompper1,lowen1,stark1,gompper2}.

A striking example can be seen in the accumulation of sperm on glass surfaces reported by Rothschild \cite{gompper1}, or in the clockwise (CW) motion that bacterium \textit{E. Coli} performs near a solid no-slip boundary, which becomes counterclockwise (CCW) near a free surface \cite{e_coli,lauga2,dileonardo1,dileonardo2}. The physical mechanism behind these behaviours are likely to be found in the chiral nature of the bacterium: the left-handed helical bundle of the flagella rotates CCW when viewed from the bundle side, with a corresponding rotation of the cell body in the CW direction. In the case of no-slip surfaces, the CW body rotation generates higher viscous stress in the gap between the cell body and the surface, resulting in a deviation to the right direction. At the same time, the flagellar bundle rotating CCW moves in the opposite direction. In the end, these combined effects determine the CW swimming of \textit{E. Coli}. J. Hu et al. modelled the motion of \textit{E. Coli} including this chiral feature and finding good agreement between numerical simulations and experiments \cite{gompper5}. In recent years, the chirality has also been experimentally and theoretically investigated in the motion of droplets of cholesteric liquid crystals \cite{sano1,sano2,sano3} and active fluids \cite{marenduzzo1,marenduzzo2}.

Another interesting example is found in the case of spherical alga called \textit{Volvox}: when two algae swim near a solid plate, they attract each other to form bound states in which they collectively move in a manner reminiscent of a dancing waltz or minuet \cite{ishikawa1}. Experimental observation of scattering motions of \textit{C. Reinhardtii} near a plate has also been reported \cite{polin1}.

Swimmers in nature are also exposed to external forces that significantly affect their motions. The gravity force is an ubiquitous example which is responsible for various phenomena including bound swimmers state \cite{ishikawa1}, rising of polar order in sedimenting swimmers \cite{stark2}, gravitaxis \cite{lowen2}, inverted sedimentation \cite{stark3}, rafting of active emulsion droplets \cite{herminghaus1}, and the formation of phytoplankton layers in coastal ocean \cite{durham}.

In this paper, we aim to numerically investigate the dynamics of swimming and sedimenting chiral microorganisms. To achieve this goal, we used the well-established squirmer model by introducing the rotlet dipole to model the chiral nature of bacteria mentioned before \cite{squirm1,squirm2,lauga3,poddar,lauga4,lauga5}. We show that depending on the type of squirmer involved, the magnitude of the chirality, and the gravity regime, different behaviours are observed. In particular, the chirality is responsible for the deviation of swimmer trajectories from a straight line.

\section{\label{sec:level1_2}SIMULATION METHODS}

\subsection{\label{sec:level2_1}The squirmer model}

%\onecolumngrid

\begin{figure}
    \begin{center}
    %\centering
        %
        \subfigure[]{%
%            \label{}
            \includegraphics[width=0.5\textwidth]{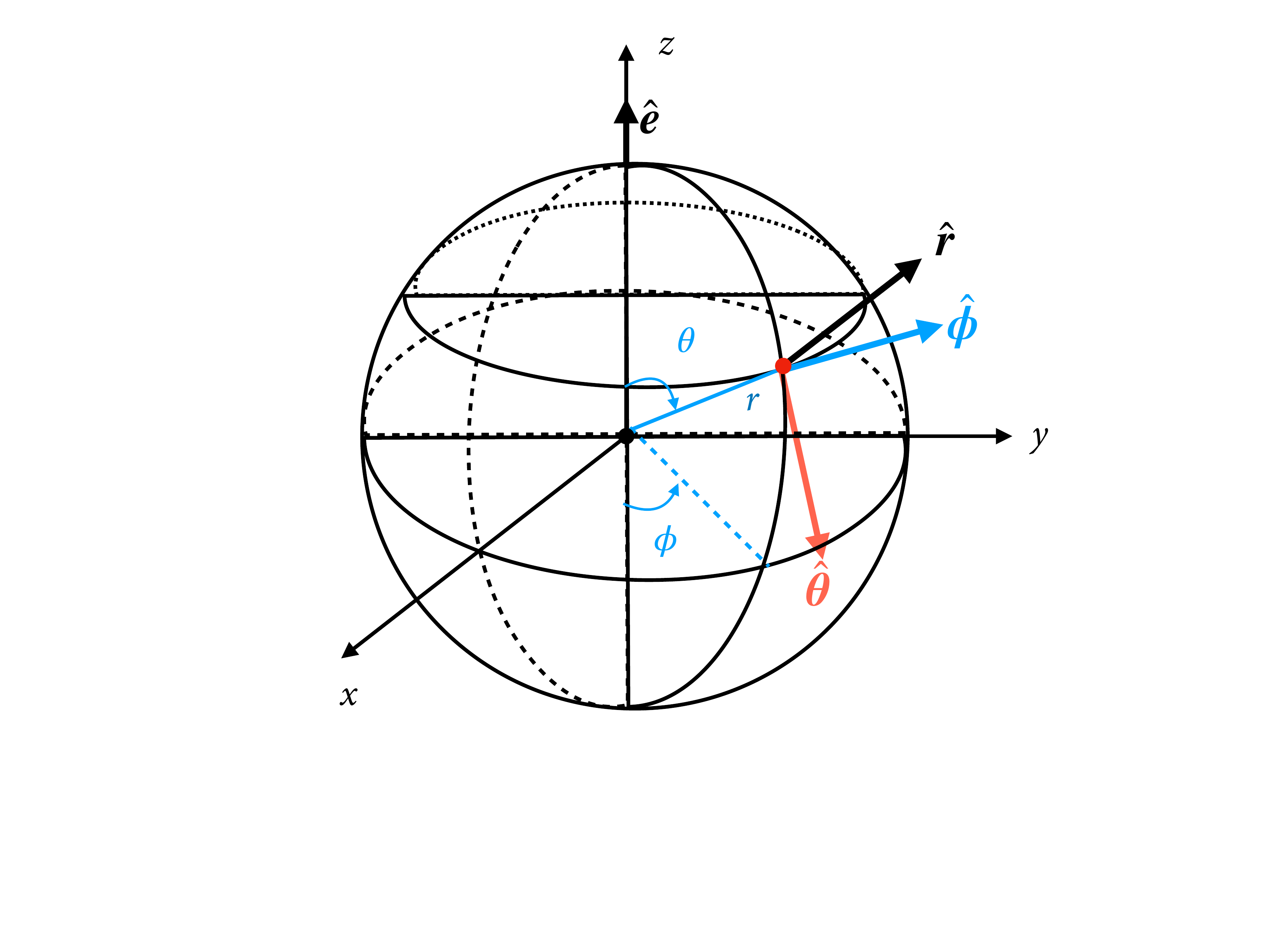}
        }        \\
        \subfigure[]{%
%            \label{}
            \includegraphics[width=0.45\textwidth]{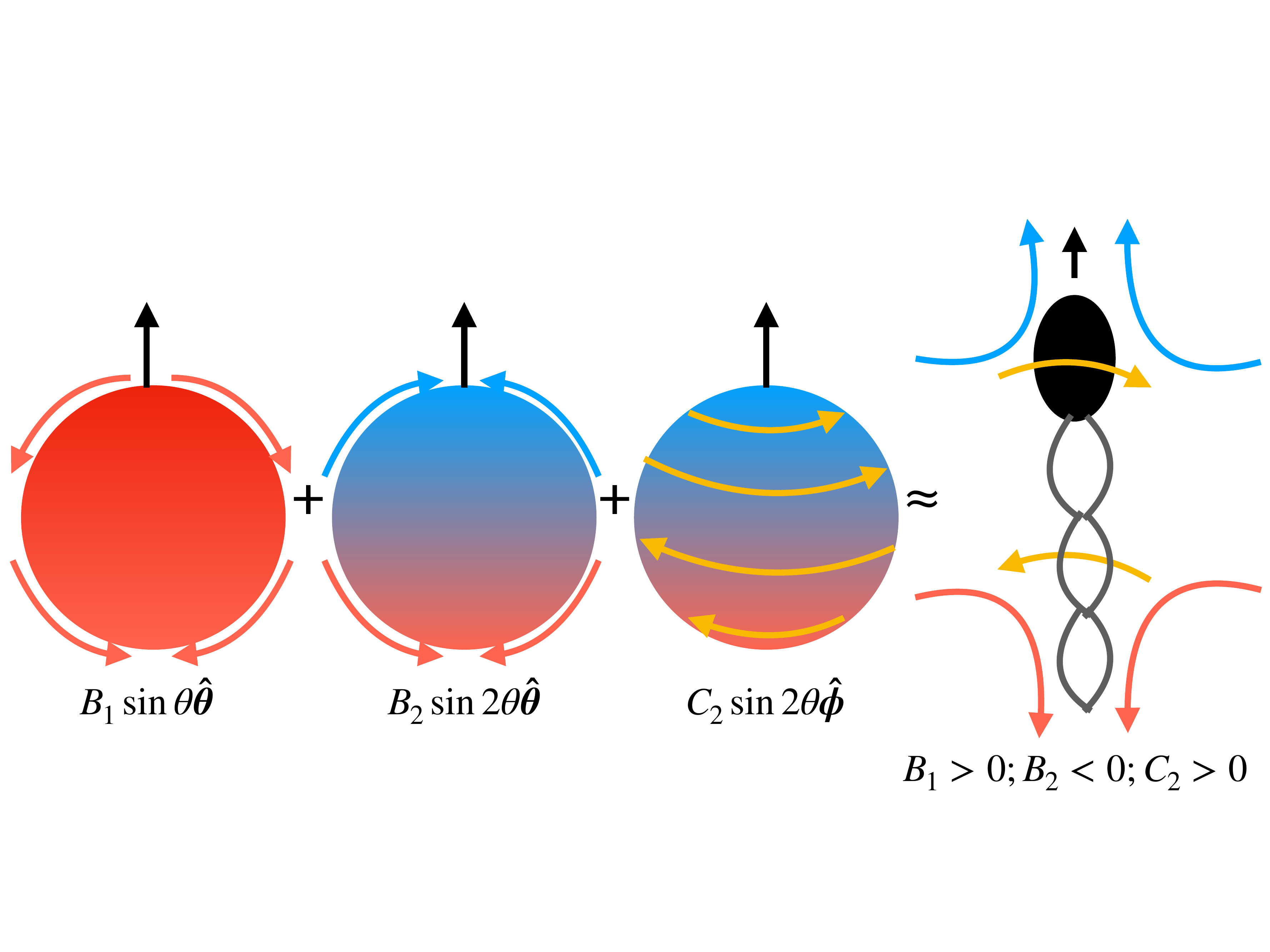}
        }

    \end{center}
    \caption{(a) Definition of the unit vectors in the spherical coordinate system: $\boldsymbol{\hat{r}}$, $\boldsymbol{\hat{\theta}}$ and $\boldsymbol{\hat{\phi}}$, with $\boldsymbol{\hat{e}}$ the swimming direction. (b) Schematic representations of the source dipole ($B_{1}\sin\theta\boldsymbol{\hat{\theta}}$), stresslet ($B_{2}\sin2\theta\boldsymbol{\hat{\theta}}$), and rotlet dipole ($C_{2}\sin\theta\boldsymbol{\hat{\phi}}$), combined to represent chiral microorganisms like a bacterium on the right.}
    \label{sketch}
\end{figure}

%\twocolumngrid
In order to model self-propelled swimmers, the squirmer model is adopted \cite{squirm1,squirm2}. It consists of a spherical object with modified stick boundary conditions, with an imposed slip velocity $\boldsymbol{u}^{s}(\hat{\boldsymbol{r}})$ at its surface, which is responsible for the self-propulsion. The most general form is given as an infinite expansion of radial, polar and azimuthal velocity components, but for simplicity the radial term is generally neglected \cite{kim,lauga6}
\begin{equation}
\begin{split}
\boldsymbol{u}^{s}(\hat{\boldsymbol{r}})&=\sum_{n=1}^{\infty}\frac{2}{n(n+1)}B_{n}P'_{n}(\cos\theta)\sin\theta\boldsymbol{\hat{\theta}}\\
& +\sum_{n=1}^{\infty}C_{n}P'_{n}(\cos\theta)\sin\theta\boldsymbol{\hat{\phi}},
\end{split}
\end{equation}
where $\boldsymbol{\hat{\theta}}$, $\boldsymbol{\hat{\phi}}$ and $\hat{\boldsymbol{r}} $ are the polar, azimuthal and radial unit vectors for a given point at the surface of the particle, $\theta=\cos^{-1}(\hat{\boldsymbol{r}} \cdot \hat{\boldsymbol{e}})$ is the polar angle and $\phi$  the azimuthal angle, with $\hat{\boldsymbol{e}}$ the swimming direction as shown in Fig.~\ref{sketch} (a).
Usually, this series expansion is truncated to second order in the polar component, and the azimuthal terms are ignored. Here we explicitly take into account the azimuthal component (to second order) which leads to the following slip velocity:
\begin{equation}
\boldsymbol{u}^{s}(\theta,\phi)=B_{1}\left ( \sin \theta +\frac{\beta}{2}\sin2\theta \right )\boldsymbol{\hat{\theta}} +\frac{3}{2}C_{2}\sin2\theta\boldsymbol{\hat{\phi}}.
\label{squirm_eq}
\end{equation}
The coefficient $B_{1}$ in Eq.~(\ref{squirm_eq}) is physically related to the steady-state swimming velocity of the squirmer $v_{0}=2/3B_{1}$, the ratio $\beta=B_{2}/B_{1}$ determines the pusher/puller type and its strength. When $\beta$ is negative, the squirmers are pushers and generate extensile flow fields along the swimming axis; when $\beta$ is positive, the squirmers are pullers generating contractile flow fields. The case of $\beta=0$ corresponds to the neutral squirmer. The main difference in the types of swimmers can be related to the position of the propulsion mechanism along the body. A pusher is a swimmer where the propulsion is generated at the back (e.g., bacteria like \textit{E. Coli}), while for pullers, the propulsion comes from the front (e.g., algae-like \textit{C. Reinhardtii}). In neutral swimmers (like \textit{Volvox}), the coefficient $B_{2}$ is small compared to $B_{1}$, and this is reflected in a symmetric flow field without vorticity. Finally, the velocity field decays as $r^{-3}$ for neutral swimmers, while it decays as $r^{-2}$ for pushers and pullers. As regards the azimuthal component of the surface velocity, the first term $C_{1}$ is the so-called rotlet, decaying as $r^{-2}$, which is neglected because it can't exist in a torque-free system \cite{pedley}. The first non-trivial coefficient $C_{2}$ is the so-called rotlet dipole, which decays as $r^{-3}$ and is physically related to the chiral nature of swimming microorganisms, like the previously mentioned \textit{E. Coli} \cite{e_coli,lauga2,poddar,lauga3,lauga4,lauga5,dileonardo1,dileonardo2}. 

In order to quantify the strength of the rotlet dipole, we define the chiral dimensionless parameter $\chi=C_{2}/B_{1}$. In the limiting case of $\chi=0$, the standard squirmer model with a sticky polar surface velocity, adopted in previous literature, is recovered. Figure \ref{sketch} (b) shows a sketch of all the polar (source dipole and stresslet) and azimuthal components (rotlet dipole) used in this study to model chiral microorganisms.

\subsection{\label{sec:level2_2}The Smoothed Profile Method}

In order to correctly solve the dynamics of the system, the coupled equations of motion for the viscous host fluid and the swimming squirmer need to be considered \cite{oyama1,oyama2,oyama3}. The squirmer model is incorporated in the \textit{Smoothed Profile Method} (SPM), a numerical technique to solve for the particle-fluid coupling with fully resolved hydrodynamics \cite{yamamoto1,yamamoto2,yamamoto3}.
The dynamics of the particle is governed by the Newton-Euler equations of motion:

\begin{align}
\begin{split}
\dot{\boldsymbol{R}}_{i}&=\boldsymbol{V}_{i}, \\
\dot{\boldsymbol{Q}}_{i}&=\text{skew}(\boldsymbol{\Omega}_{i})\cdot \boldsymbol{Q}_{i},\\
M_{p}\dot{\boldsymbol{V}}_{i}  &=\boldsymbol{F}_{i}^{H}+\boldsymbol{F}_{i}^{C}+\boldsymbol{F}_{i}^{ext}, \\ 
\boldsymbol{I}_{p} \cdot \dot{\boldsymbol{\Omega}}_{i}&=\boldsymbol{N}_{i}^{H} + \boldsymbol{N}_{i}^{ext},
\end{split}
\numberthis
\label{particle}
\end{align}
where $i$ is the particle index, $\boldsymbol{R}_{i}$ and $\boldsymbol{V}_{i}$ are the center of mass position and velocity, $\boldsymbol{Q}_{i}$ is the orientation matrix, skew$(\boldsymbol{\Omega}_{i})$ is the skew symmetric matrix of the angular velocity $\boldsymbol{\Omega}_{i}$, $\boldsymbol{I}_{p}(=2/5M_{p}R^{2}\boldsymbol{I})$ is the inertia tensor (with $\boldsymbol{I}$ is the unit tensor), and $M_{p}(=\frac{4}{3}\pi R^{3}\rho_{p})$ is the mass of a spherical particle with density $\rho_{p}$ and radius $R$. $\boldsymbol{F}_{i}^{H}$ ($\boldsymbol{N}_{i}^{H}$) is the force (torque) which arises from the hydrodynamic flow field around the swimmer, $\boldsymbol{F}_{i}^{C}$ is the particle-particle force due to the steric repulsion, and $\boldsymbol{F}_{i}^{ext}$ ($\boldsymbol{N}_{i}^{ext}$) is the external force (torque). In this study, the gravitational force $\boldsymbol{F}_{i}^{ext}=-\frac{4}{3}\pi R^{3}(\rho_{p}-\rho_{f})g\boldsymbol{z}$ is applied that causes the squirmer to sediment along the vertical $z$ axis with $g$ the acceleration gravity and $\rho_{f}$ density of the fluid \cite{yamamoto4,yamamoto5,yamamoto6,yamamoto7,yamamoto8}. The center mass of the particle is assumed to coincide with its geometric center, however, more general cases in which they are displaced ({\it i.e.}, bottom-heaviness) would result in an additional torque to be taken into account \cite{ishikawa2,ishikawa3,ishikawa4,ishikawa5,ishikawa7,ishikawa6}.

The evolution of the host fluid is governed by the Navier-Stokes equation with the incompressible condition:
 \begin{equation}
 \nabla \cdot \boldsymbol{u}_{f}=0,
 \label{fluid1}
  \end{equation}
\begin{equation}
 \rho_{f}(\partial_{t}+\boldsymbol{u}_{f} \cdot \nabla)\boldsymbol{u}_{f}=\nabla \cdot \boldsymbol{\sigma}_{f}, 
 \label{fluid2}
 \end{equation}
 \begin{equation}
 \boldsymbol{\sigma}_{f}=-p\boldsymbol{I}+\eta_{f}\left \{\nabla \boldsymbol{u}_{f} +(\nabla \boldsymbol{u}_{f})^{T} \right \},
  \end{equation} 
  where $\boldsymbol{u}_{f}$ is the fluid mass density, $\eta_{f}$ the shear viscosity, and $\boldsymbol{\sigma}_{f}$ is the Newtonian stress tensor.

 The key element of the SPM, to solve Eqs.~(\ref{particle})-(\ref{fluid2}), is to replace the sharp boundaries between the solid particles and the host fluid with interfacial regions with a finite width $\xi_{p}$. For this, we introduce a smooth continuous function $\phi_{p}$ taking values $1$ in the solid domain and 0 in the fluid domain. In this way, it is possible to define the total velocity $\boldsymbol{u}$, which satisfies the correct momentum exchange between the solid particles and the host fluid, as
\begin{equation}
\boldsymbol{u}=(1-\phi_{p})\boldsymbol{u}_{f}+\phi_{p}\boldsymbol{u}_{p}+\phi_{W}\boldsymbol{u}_{W},
\end{equation}
 where $(1-\phi_{p})\boldsymbol{u}_{f}$ is the contribution from the fluid, 
\begin{equation}
\phi_{p}\boldsymbol{u}_{p}=\sum_{i}\phi_{p}[\boldsymbol{V}_{i}+\boldsymbol{\Omega_{i}} \times \boldsymbol{R}_{i}]
\end{equation}
is the contributions from the rigid particles, and $\phi_{W}\boldsymbol{u}_{W}$ arises from the non-moving flat plates placed at the top and the bottom of the system, normal to the vertical $z$ direction. In implementing the rigid flat plates, we introduced an additional phase field function $\phi_{W}$ which connects the values $1$ in the plate domain and $0$ outside with a finite width $\xi_{W}$. The plate velocity is defined as $\boldsymbol{u}_{W}=0$. The time evolution of the total flow field $\boldsymbol{u}$ then obeys
 \begin{equation}
\nabla \cdot \boldsymbol{u}=0,
\end{equation}
\begin{equation}
 \rho_{f}(\partial_{t}+\boldsymbol{u} \cdot \nabla)\boldsymbol{u}=\nabla \cdot \boldsymbol{\sigma}_{f} +\rho_{f}(\phi_{p} \boldsymbol{f}_{p}+\boldsymbol{f}_{sq}),
\end{equation}
 where $\phi_{p} \boldsymbol{f}_{p}$ is the body force necessary to maintain the rigidity of the particles, and $\boldsymbol{f}_{sq}$ is the force due to the active squirming motion.

\section{\label{sec:level1_3}RESULTS}

\subsection{Sedimentation without chirality} 

\begin{figure}
    \begin{center}
        \subfigure[]{%
%            \label{}
            \includegraphics[width=0.37\textwidth]{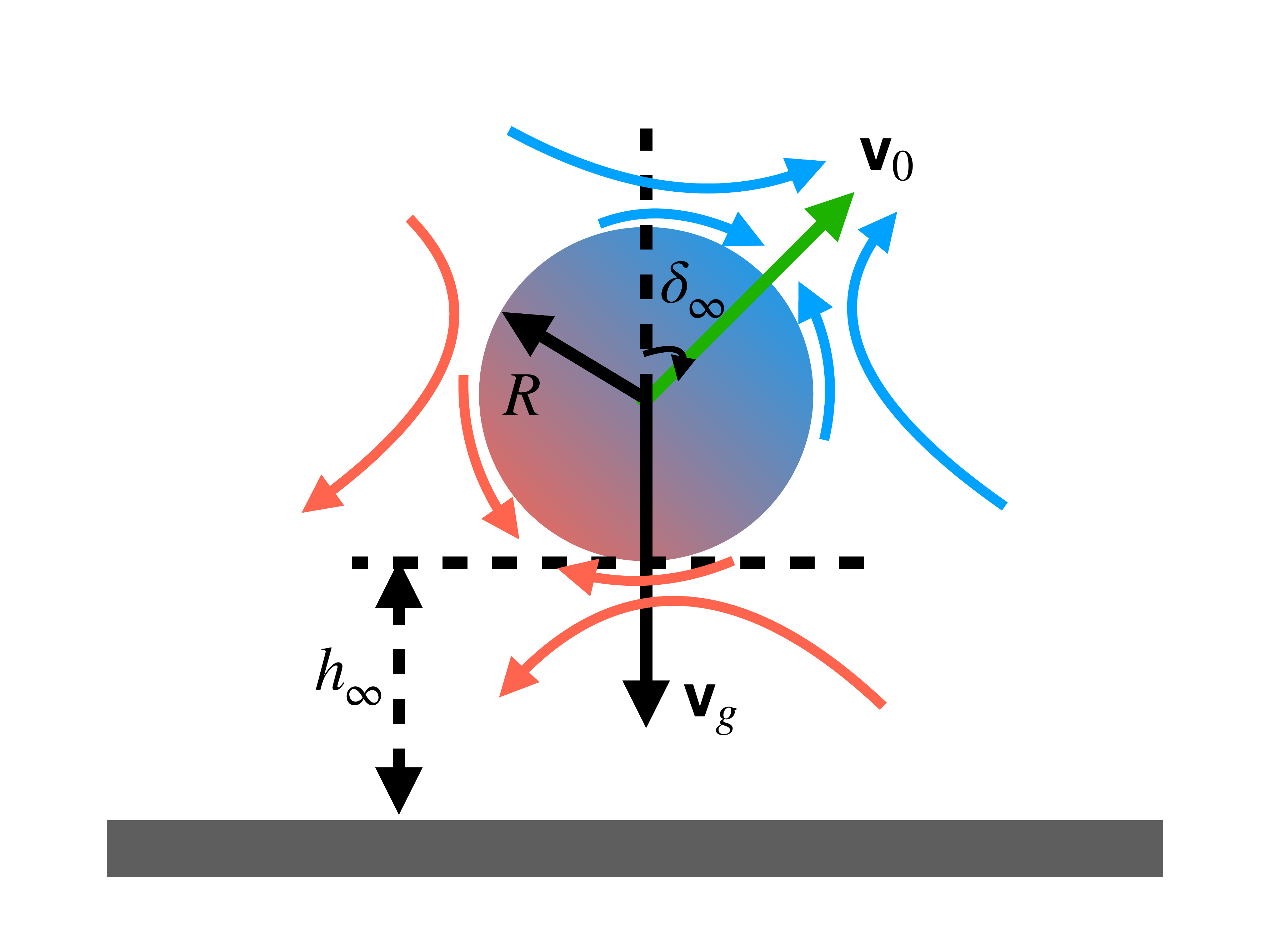}
        }                \\
        \subfigure[]{%
%            \label{}
            \includegraphics[width=0.35\textwidth]{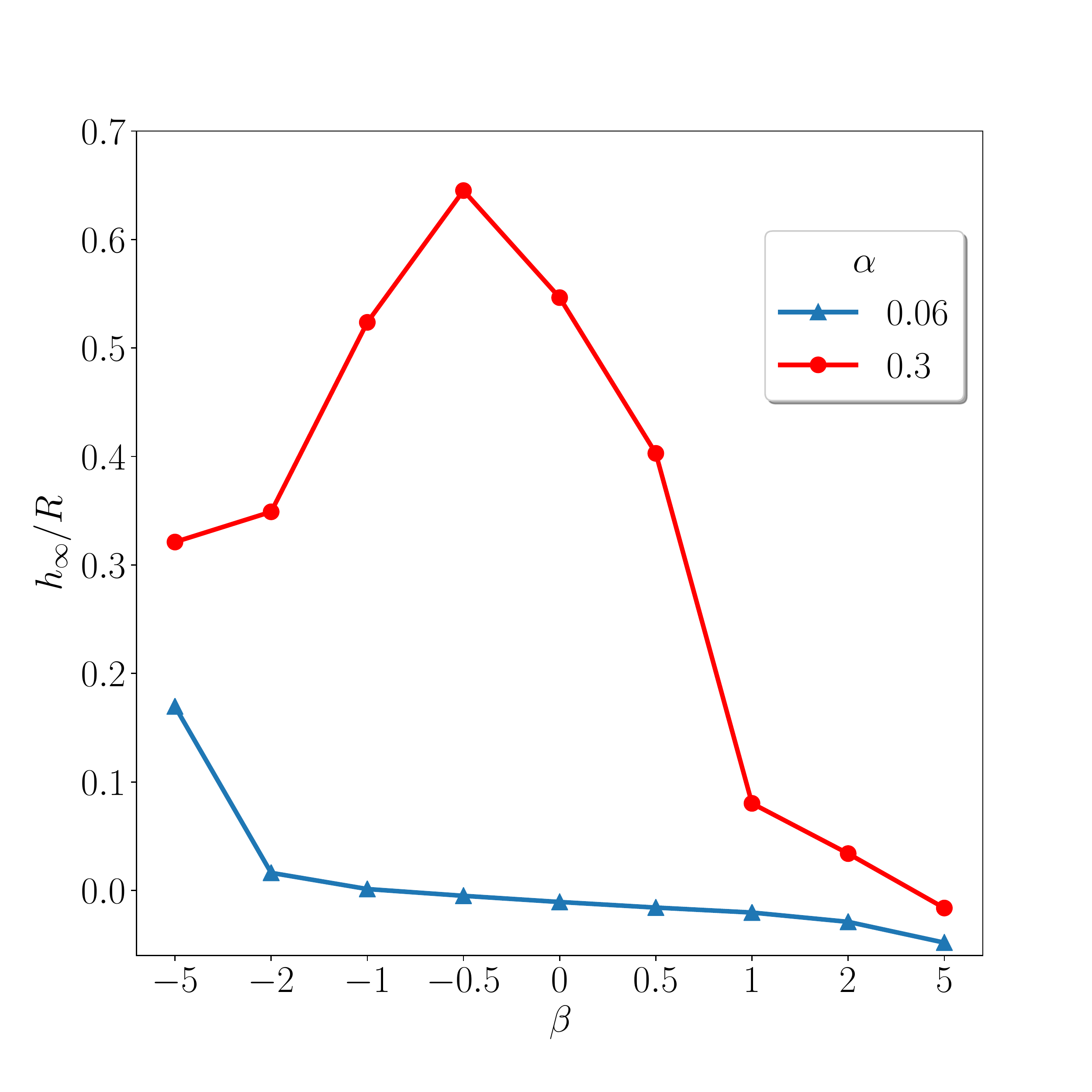}
        }                \\
        \subfigure[]{%
 %           \label{}
            \includegraphics[width=0.35\textwidth]{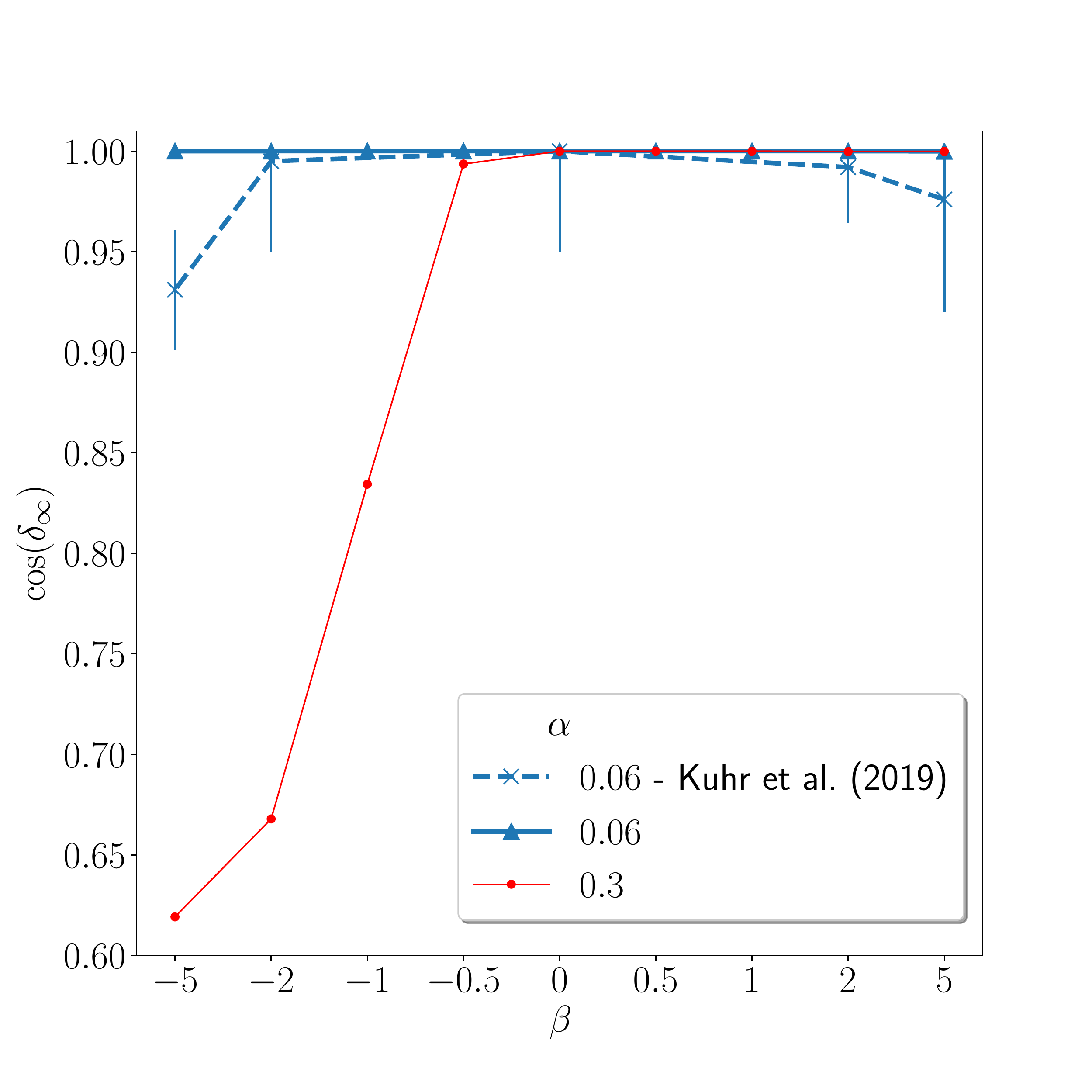}
        }
    \end{center}
    \caption{(a) A schematic representation of a micro-swimmer (pusher with radius $R$) on the bottom plate, and the definitions of the height $h_{\infty}$, the single particle swimming velocity $v_{0}$, oriented an angle $\delta_{\infty}$ from the plate normal axis, with sedimenting velocity $v_{g}$. (b) The stationary height of a micro-swimmer at various $\beta$, under moderate $\alpha=0.3$ (red curve) and strong $\alpha=0.06$ (blue curve) gravity, without chirality $\chi=0$. (c) The stationary orientation of microswimmers with the same parameters of (a). The data for $\alpha=0.06$ from the work of Kuhr et al.~\cite{stark6} (blue dashed curve) are also shown, with the error bars indicating the magnitudes of the thermal fluctuations associated with the MPCD simulations used in their study.}
    \label{no_chirality}
\end{figure}

In the first analysis, we studied the sedimentation of a single squirmer under gravity near a solid flat plate with no chirality $\chi=0$. 
It is useful to define two dimensionless parameters to characterize the dynamics. 
The first one is the Reynolds number $Re=\rho_{f} v_{0} 2R/\eta_{f}$. 
We have considered $Re=0.1$ which is an acceptable value to fullfill the constraint of $Re \ll 1$ for swimming micro-organisms \cite{lauga1,marchetti1,gompper1,lowen1,stark1,gompper2}. 
We chose values of the radius and the interface width of the particle and wall of $R=5\Delta$, $\xi_{p}=2\Delta$, $\xi_{W}=4\Delta$ and respectively, where $\Delta=1$ is the grid spacing. 
The shear viscosity $\eta_{f}$ and the fluid density $\rho_{f}$ are also set to $1$. 
The second dimensionless number represents the strength of sedimentation, which accounts for the effect of gravity compared with the self-propulsion, defined as $\alpha=v_{0}/v_{g}$, the ratio between the natural self-propelling velocity of the squirmer $v_{0}=2/3B_{1}$ and the sedimenting velocity of a corresponding passive particle $v_{g}=M_{p}g/6 \pi \eta_{f} R$ \cite{stark4,stark5,stark6,lintuvuori1,lintuvuori2}.
If $\alpha \geq 1$, the self-propulsion dominates over the sedimentation due to gravity; we refer to this state as a cruising regime. 
If $\alpha \ll 1$, the squirmer sediments to the bottom, reorients perpendicular to the plate, and finally stops due to the strong gravity; we refer to this state as the strong gravity regime. 
Therefore, in order to observe non-trivial dynamical states, we chose $\alpha=0.3$ which is in a moderate gravity regime in addition to $\alpha=0.06$ which is in the strong gravity regime. 
To ensure that the squirmer sediments under gravity, it was assumed a density ratio of $\rho_{p}/\rho_{f}=5$. 
In this case, the system consists of a cubic box of $L\times L\times L$, with $L=64\Delta$. 
In the simulations, we placed a single squirmer in the initial position $(L/2\Delta, L/2\Delta, L/7\Delta)$, near the bottom plate, with its swimming axis in the $y-$ direction parallel to the bottom plate, and let the system evolve until reaching to a stationary state. 
We also performed a set of simulations starting with different initial orientations, and confirmed that the squirmer always reaches the same stationary state for the range of parameter values $\beta \in [-5, 5]$ and $\chi \in [-5, 5]$ considered in the present study.

Figure \ref{no_chirality} (b) shows the surface to surface distance  $h_\infty$ between the squirmer and the bottom plate, defined schematically in Fig. \ref{no_chirality} (a), for various values of $\beta$, in the moderate ($\alpha=0.3$) and strong ($\alpha=0.06$) gravity regimes. 
The competition between sedimentation and self-propulsion results in a  non-zero separation from the bottom plate. For $\alpha=0.06$, the gravity force is strong enough to push the squirmer down to the bottom plate, $h_\infty/R<0.2$. For $\alpha=0.3$, instead, highly asymmetric dynamics for both pushers ($\beta<0$) and pullers ($\beta>0$) are observed, with the stationary height much higher than that of the strong gravity case. 

Figure \ref{no_chirality} (c) shows the stationary orientations of the squirmer $\cos(\delta_\infty)$ at various values of $\beta$. 
The stationary angle $\delta_\infty$ is defined as the angle between the swimming direction and the normal direction to the plate, as shown in Fig. \ref{no_chirality} (a). 
In case of strong gravity $\alpha=0.06$, the squirmers sediment down to the bottom plate and reorient in perpendicular direction corresponding to $\cos(\delta_{\infty})\simeq 1$. 
In the case of moderate gravity $\alpha=0.3$, pushers move in tilted directions, while neutral swimmers and pullers still reorient perpendicular to the plate (see movies S1-S3 of the Supplementary Material).

The fact that pushers in Fig. \ref{no_chirality} (b) can reach stationary heights greater than the correspondent pullers is due to their nature of pushing fluid away at the back. 
In particular, the maximum in the stationary height for $\alpha=0.3$,  around the weak pusher condition $\beta \simeq -0.5$, is related to the stationary orientation shown in Fig. \ref{no_chirality} (c).  The pushers, in contrast to the pullers, tend to develop a tilted orientation for $\beta\le -0.5$, as also predicted by the lubrication theory and far field-approximations \cite{stark4,stark6,lintuvuori3}.

The data for $\alpha=0.06$ from the work of Kuhr et al.~\cite{stark6} are also shown in Fig. \ref{no_chirality} (c), with the error bars indicating the magnitudes of the thermal fluctuations associated with the multi-particle collision dynamics (MPCD) used in their simulations.
Excellent agreement is seen between our results and the date of Kuhr et al. of \cite{stark6} for $-2\le\beta\le2$, where the squirmers orient perpendicular to the bottom plate due to the strong gravity.  
However, non-negligible deviations emerge for $|\beta|>0.5$ while our results for $|\beta|<2$ still agree with Kuhr et al. within the error bars.
Possible explanations are the following: 1) The discretization error may be non-negligible because we solve the governing equations on a fixed Cartesian grid with a finite spacing $\Delta$. 
Although this type of error increases with increasing $|\beta|$, we confirmed that the discretization error remains small with the present resolution $R/\Delta=5$ in the parameter range $|\beta|\le5$ \cite{yamamoto3}. 
2) The influence of thermal fluctuation, which can not be excluded from the MPCD used in the work of Kuhr et al.~\cite{stark6}.

\subsection{Sedimentation with chirality} 

\begin{figure}
    \begin{center}
        \subfigure[]{%
%            \label{}
            \includegraphics[width=0.38\textwidth]{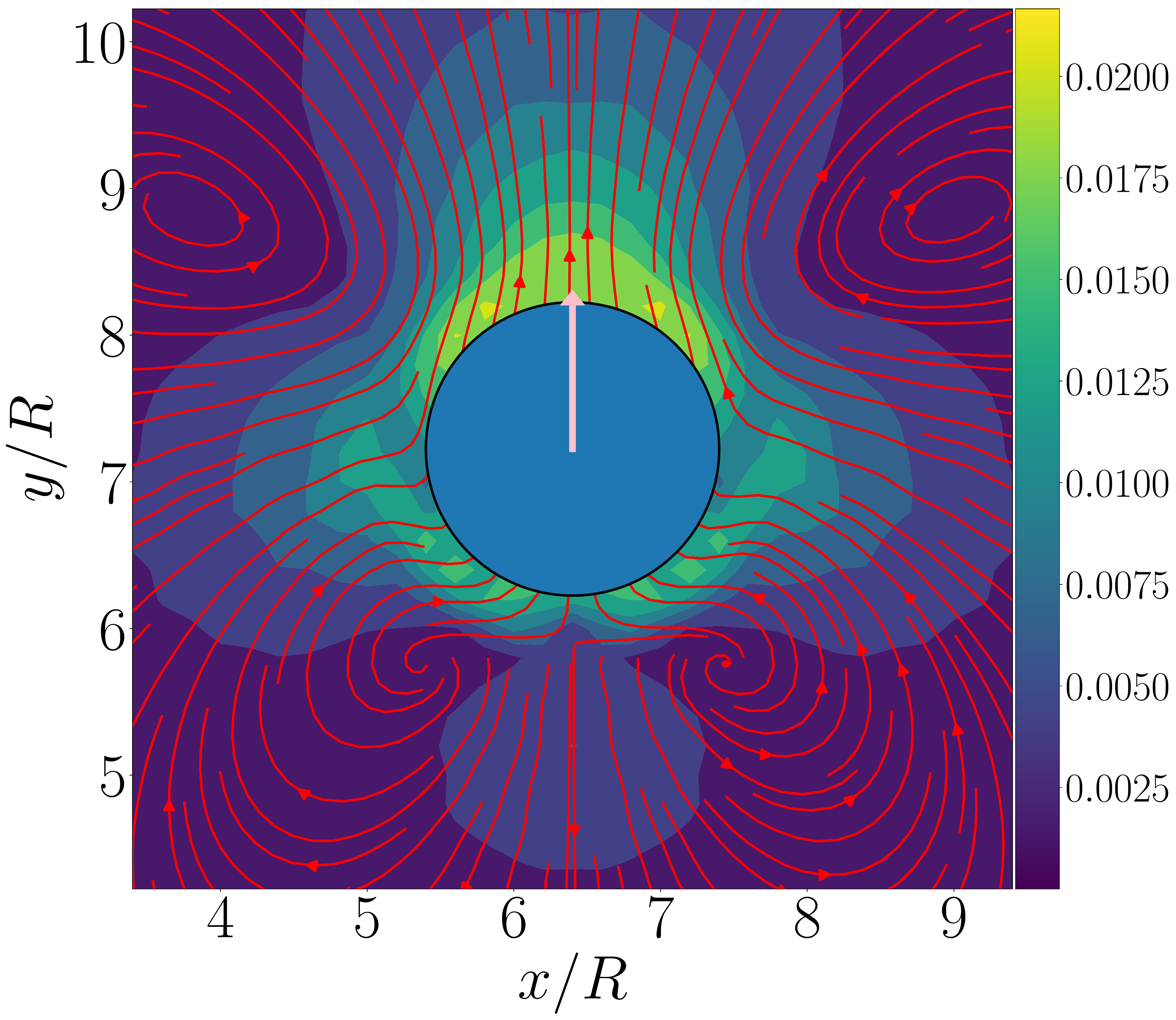}
        }                \\
        \subfigure[]{%
%            \label{}
            \includegraphics[width=0.38\textwidth]{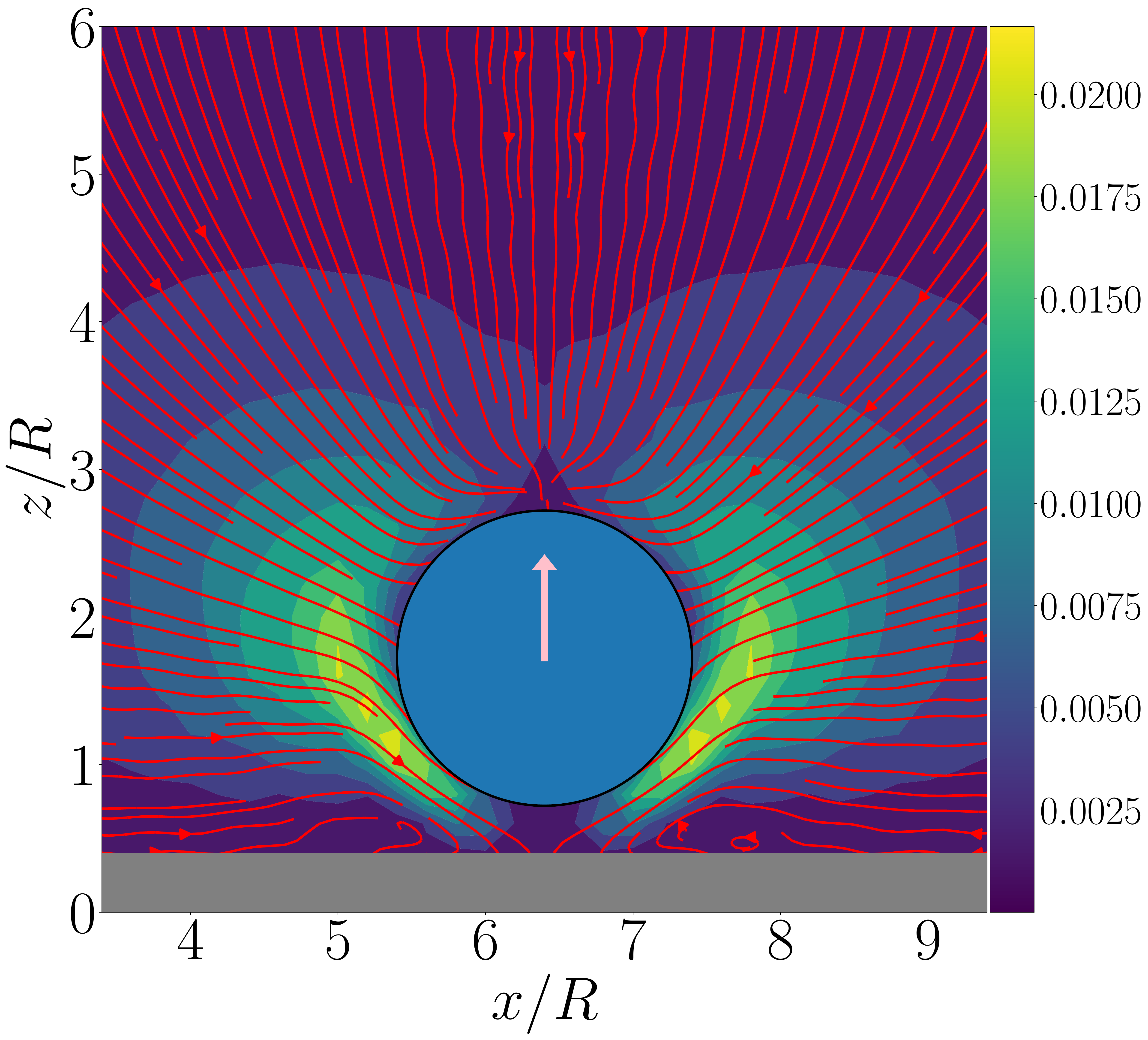}
        }                \\
        \subfigure[]{%
%            \label{}
            \includegraphics[width=0.38\textwidth]{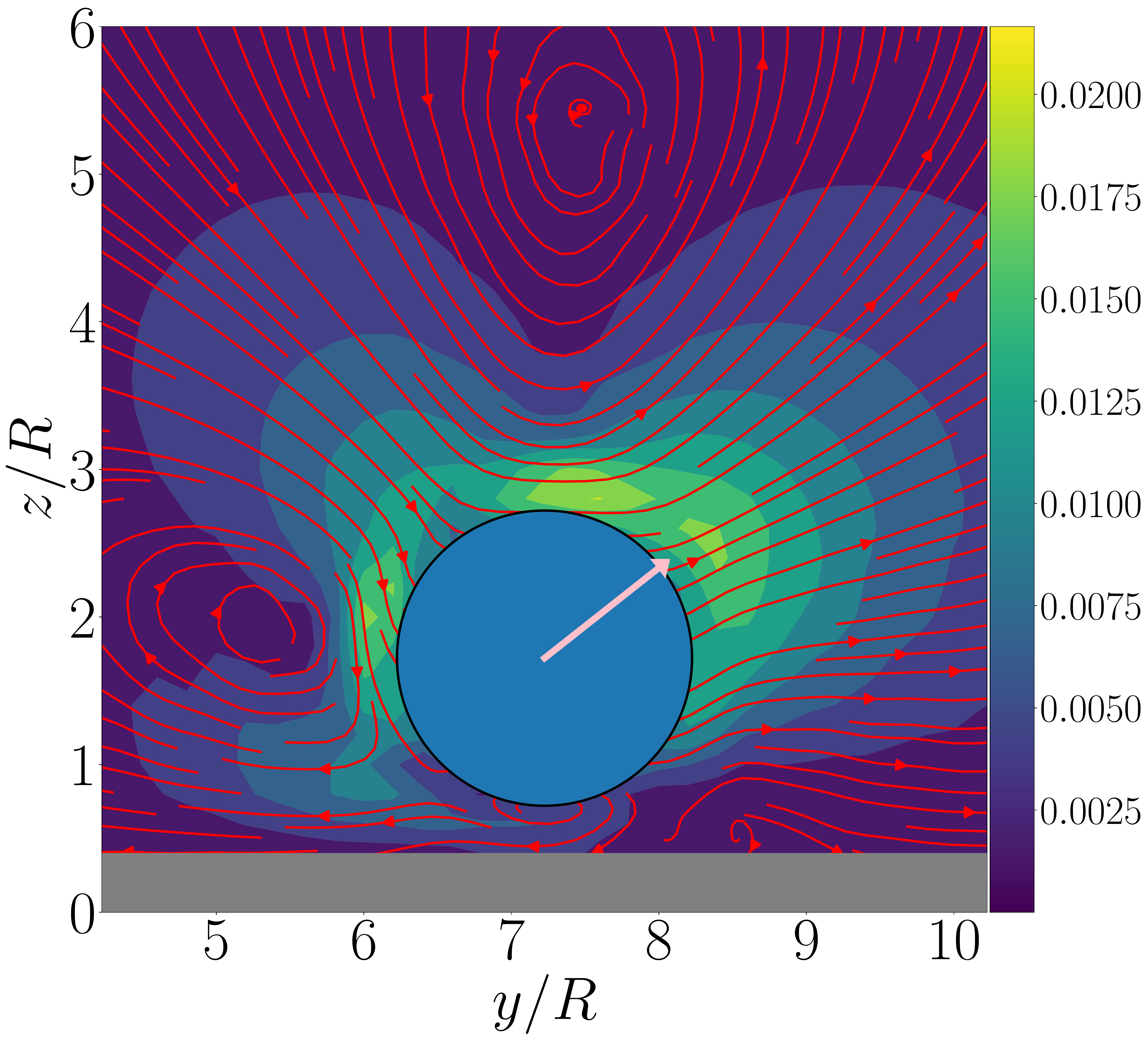}
        }
    \end{center}
    \caption{The velocity streamlines with superimposed contour plot of its magnitude around a pusher particle with $\beta=-5$ for $\chi=0$ and $\alpha=0.3$ in the (a) $xy$, (b) $xz$, and (c) $yz$ cross-sections. The pink arrows mark the swimming axis of the squirmer. The grey rectangle boxes represent the bottom plate.}
    \label{vel_no_chirality}
\end{figure}

\begin{figure}
    \begin{center}
        \subfigure[]{%
%            \label{}
            \includegraphics[width=0.37\textwidth]{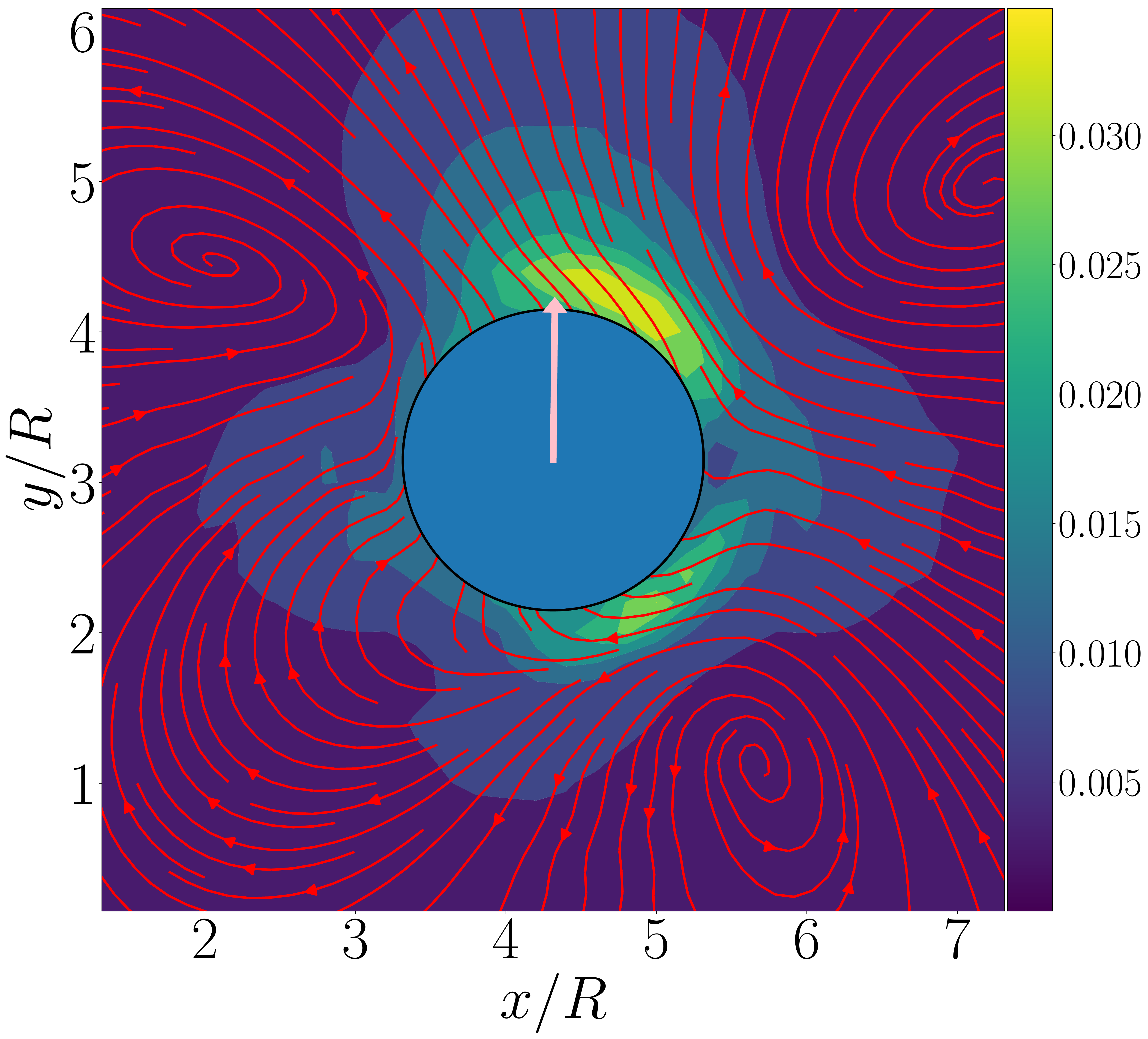}
        }                \\
        \subfigure[]{%
%            \label{}
            \includegraphics[width=0.37\textwidth]{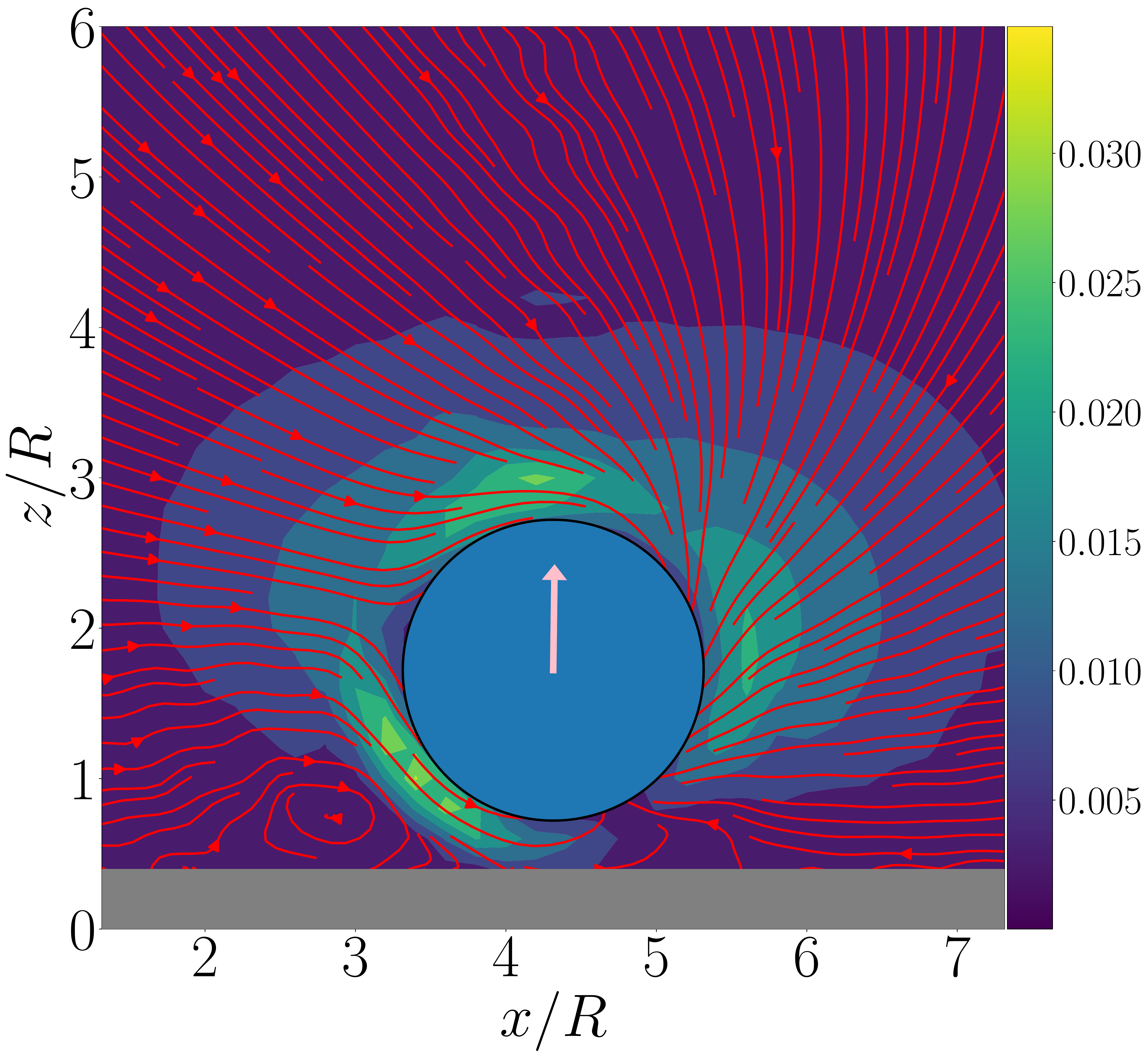}
        }                \\
        \subfigure[]{%
%            \label{}
            \includegraphics[width=0.37\textwidth]{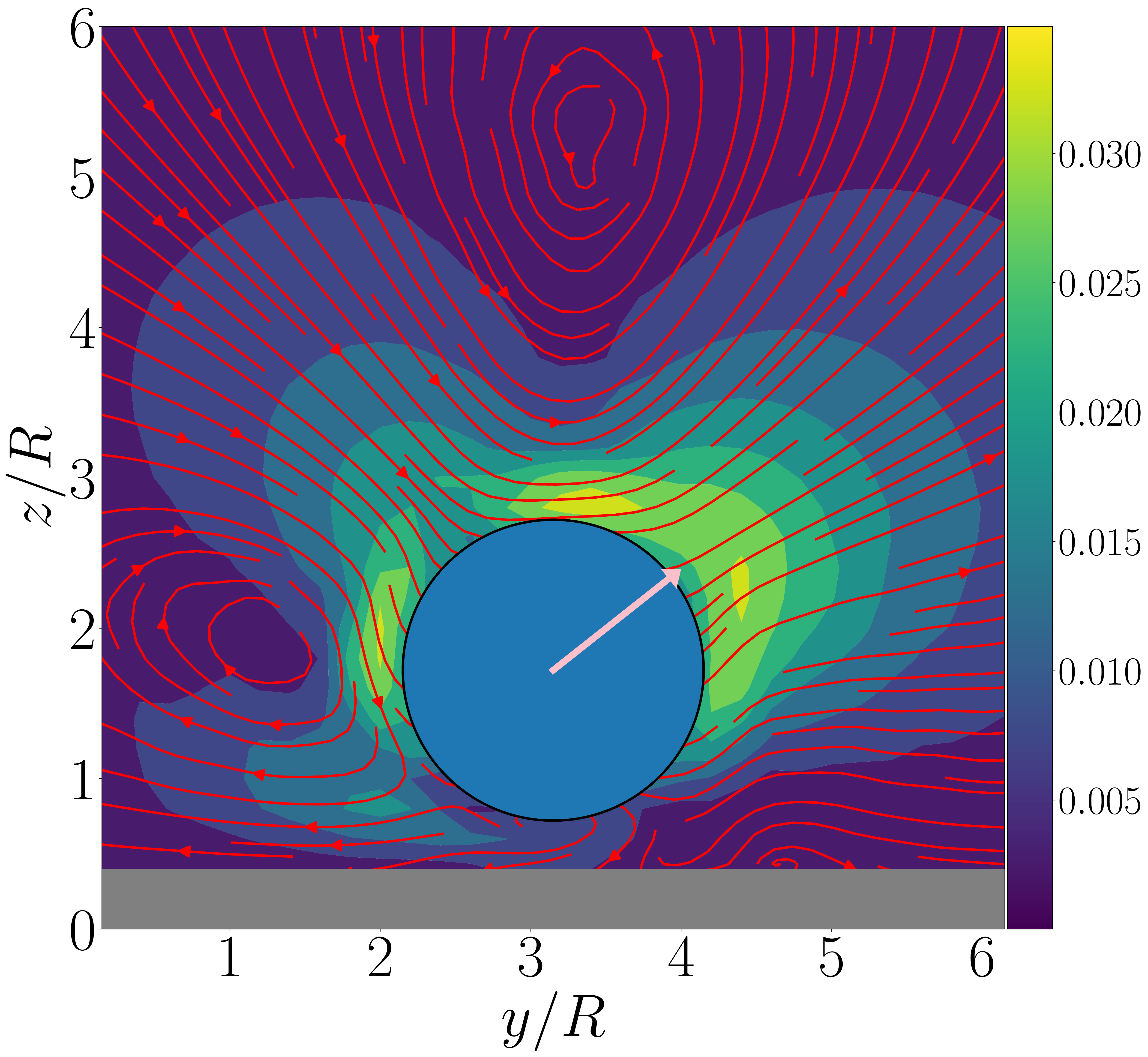}
        }
    \end{center}
    \caption{The velocity streamlines with superimposed contour plot of its magnitude around a pusher particle with $\beta=-5$ for $\chi=1$ and $\alpha=0.3$ in the (a) $xy$, (b) $xz$, and (c) $yz$ cross-sections. The pink arrows mark the swimming axis of the squirmer. The grey rectangle boxes represent the bottom plate.}
    \label{vel_chirality}
\end{figure}

The next step consisted of adding the chirality to study sedimentation under gravity near a solid flat plate. This is achieved via the term $\frac{3}{2}C_{2}\sin2\theta\boldsymbol{\hat{\phi}}$ in Eq. (\ref{squirm_eq}), corresponding to the rotlet dipole. For this analysis, we only considered the case $\alpha=0.3$.
In Fig. \ref{vel_no_chirality}, the velocity streamlines, with contour plot of its magnitude, around a pusher particle with $\beta=-5$ for $\chi=0$,  is shown in the (a) $xy$, (b) $xz$, and (c) $yz$ cross-sections. The pink arrows mark the swimming axis of the squirmer, and the grey boxe represents the bottom plate.
Fig. \ref{vel_chirality} shows the same plots for a squirmer with chirality $\chi=1$. 
In both Fig. \ref{vel_no_chirality} (c) and \ref{vel_chirality} (c), the characteristic pusher velocity fields are recognisable with fluid pushed away, along the direction of the swimming axis, and attracted in the perpendicular direction. 
A velocity vortex is formed at the back. 
The differences in the velocity fields are visible in the $xy$ and $xz$ planes, Fig. \ref{vel_no_chirality}(a)-(b) and \ref{vel_chirality} (a)-(b). 
In Fig. \ref{vel_chirality} (a) the velocity direction in the front of the squirmer is not straight as its counterpart of Fig. \ref{vel_no_chirality} (a). 
This result is a clear sign of the fact that the trajectory of the chiral pusher, in the $xy$ plane, is not straight as the one without chirality. 
Also, the plots (b) in Fig. \ref{vel_no_chirality} and \ref{vel_chirality} differ from each other, with the velocity in \ref{vel_chirality} (b) showing a vortex structure due to the interaction of the rotlet dipole with the plate.

\begin{figure}
  \includegraphics[width=\linewidth]{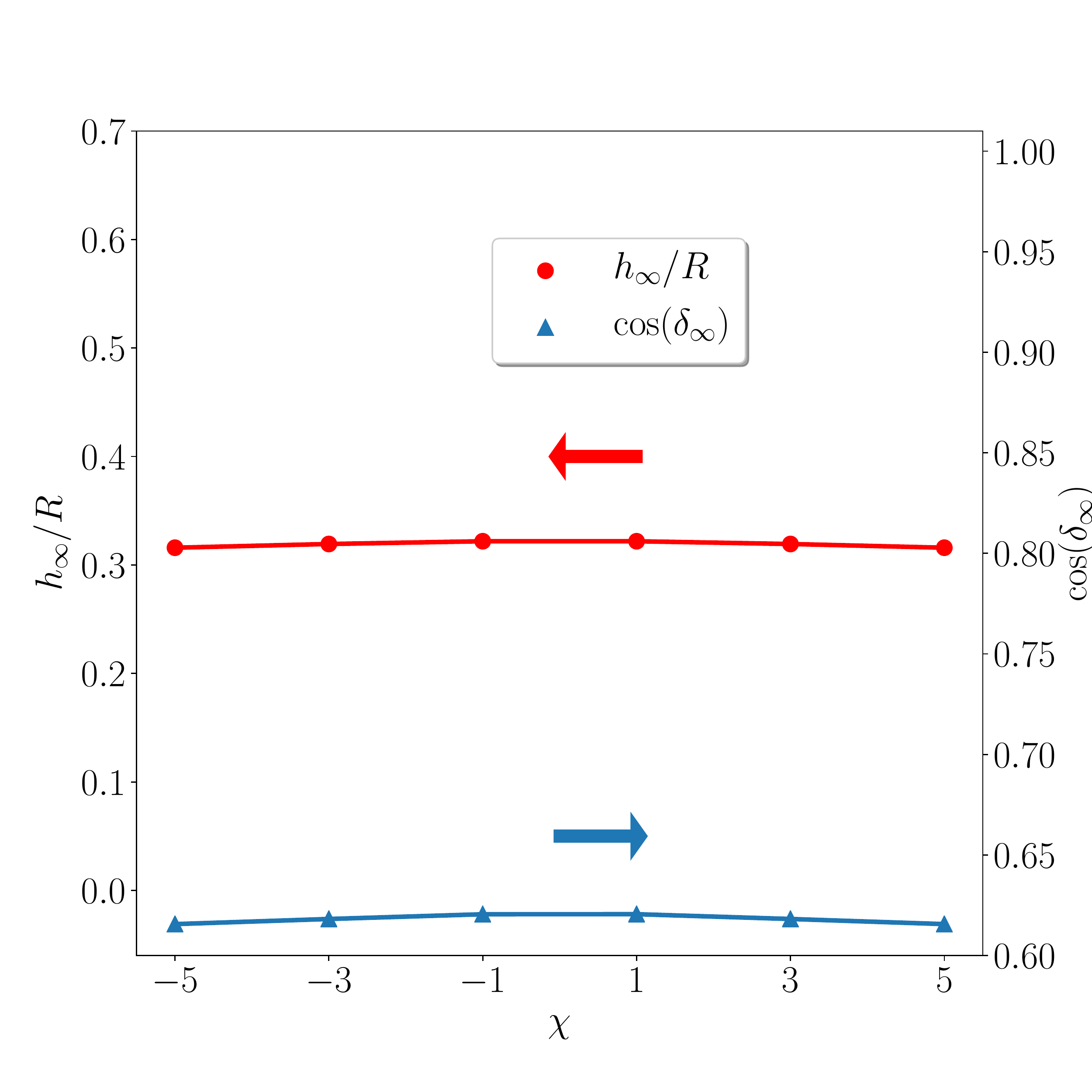}
  \caption{The stationary height (red) and orientation (blue) for a pusher particle with $\beta=-5$ at various chiralities $-5\le\chi\le5$ under moderate gravity $\alpha=0.3$.}
  \label{chirality_constant}
\end{figure}

We repeated the same analyses of the previous paragraph, evaluating the heights and orientations of the various squirmers, and we found that the introduction of the chirality did not contribute to a dramatic change in these quantities. 
This tendency is visible in Fig. \ref{chirality_constant}, where the stationary heights (left side) and orientations (right side) for the pusher $\beta=-5$ at various $\chi$ values are shown.

\begin{figure}
    \begin{center}
    %\centering
        %
        \subfigure[]{%
%            \label{}
            \includegraphics[width=0.44\textwidth]{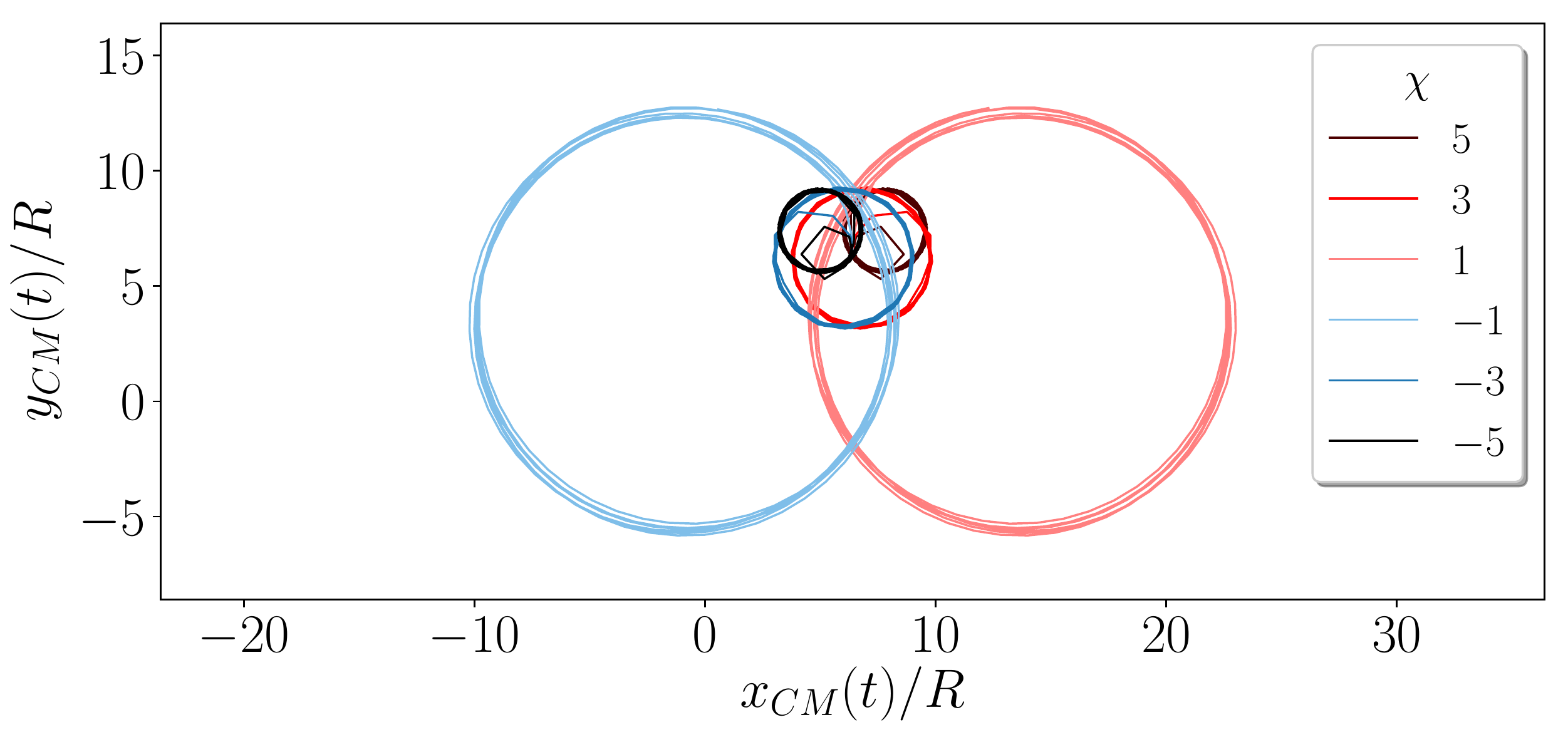}
        }        \\
        \subfigure[]{%
%            \label{}
            \includegraphics[width=0.5\textwidth]{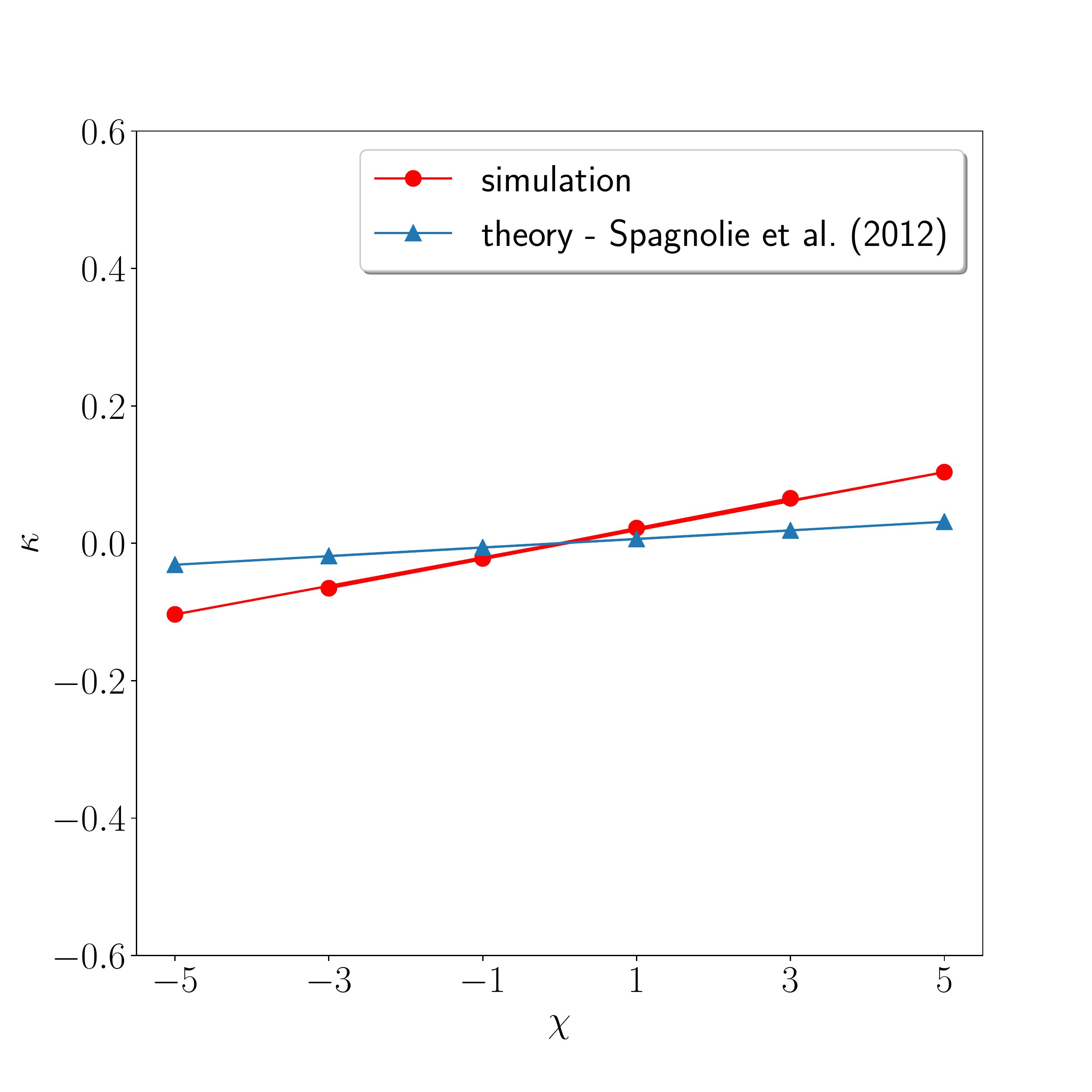}
        }

    \end{center}
    \caption{(a) The trajectories of a pusher particle $\beta=-5$ at various chiralities $-5\le\chi\le5$ under moderate gravity $\alpha=0.3$. (b) The curvature of the circular trajectories shown in (a) versus the chirality $\chi$. A theoretical expression, $\kappa=\frac{3\chi R^{3}}{32H^{4}}$ found in \cite{lauga4} is also plotted for comparison.}
    \label{chirality_plots}
\end{figure}

The physical effect of the chirality consists of a deviation of all the particles from a straight trajectory, observed without chirality, into a circular one. 
Figure \ref{chirality_plots} (a) shows the trajectories of the pusher $\beta=-5$ at various chiralities. 
For $\chi>0$, the squirmer performs CW circular motions, instead for $\chi<0$, the trajectories become CCW. 
Increasing the absolute value of $\chi$ reduces the radius of the circular trajectories of the pusher. 
We evaluated the radius of curvature according to \cite{dorst}
\begin{equation}
R_{c}=\frac{(\dot{x_{CM}}^2+\dot{y_{CM}}^2)^{3/2}}{\dot{x_{CM}}\ddot{y_{CM}}-\dot{y_{CM}}\ddot{x_{CM}}}.
\end{equation}
where the dots denote the temporal derivative, and the curvature is defined as $\kappa=R_c^{-1}$.  
If $\chi=0$, the squirmer trajectory is always straight, thus $\kappa=0$.

Figure \ref{chirality_plots} (b) shows the curvature $\kappa$ of the pusher $\beta=-5$ at various chiralities $\chi$ together with the theoretical prediction, $\kappa=\frac{3\chi R^{3}}{32H^{4}}$, proposed by Spagnolie et al. in \cite{lauga4}, Lopez et al. \cite{lauga5} and Papavassiliou et al. \cite{alexander1,alexander2}. 

Both curves show a linear trend, which suggests an increase in the curvature at increasing chirality (also shown in movies S4-S6 of the Supplementary Material). The difference between the simulated and theoretical lines in Fig. \ref{chirality_plots} (b) could be explained by taking into account three key factors. 
First of all, the theoretical expression assumes that the squirmer orients and swim parallel to the plate at a fixed distance $H$. 
This is not, our case, in which the sedimenting squirmer is parallel to the plate only at the beginning of the simulation; in the end, it acquires a perpendicular or a tilted orientation. 
Moreover, the distance from the plate is not constant during the evolution but varies as a consequence of the competition between sedimentation and hydrodynamics, reaching a final stationary value.

\section{\label{sec:level1_4}DISCUSSION}

%Various previous studies focused on the dynamics of squirmers near a plate.  Li et al., for example, analyzed these dynamics both for a single and for a collection of squirmers in inertial regime discovering different swimming mechanisms and collective structures \cite{ardekhani}.  Lintuvuori et al., instead, analyzed the dynamics of strong pushers and pullers near a plate combining both theory simulations \cite{lintuvuori3}.  In this case, it was demonstrated that the interplay between hydrodynamics and short-range repulsion could lead to trapping, periodic oscillations, and swimming speeds different from those in bulk.  In the recent work of Das et al., curved and structured plates have been considered affecting the incident angle of nearby squirmer leading to broad angular distributions  \cite{cacciuto}. 

Recent works have considered sedimenting squirmers \cite{stark4,stark5,stark6,lintuvuori1,lintuvuori2}; however, the role of chirality has not been considered yet. 
In the present study, we aimed to generalize the dynamics of a sedimenting squirmer in similar situations, but with chirality.
Shen et al. found that a neutral squirmer, weak pullers, and weak pushers reorient in the direction perpendicular to the bottom plate in \cite{lintuvuori1}, consistent with the present results shown in Fig. \ref{no_chirality} (c). 
The stationary height for pushers is higher than that for pullers and neutral squirmers because the fluid flows are stronger for pushers at their back.
These results are also consistent with R{\"u}hle et al. who considered the dynamics of a single squirmer without chirality under gravity with various values of $\alpha$ and $\beta$ in \cite{stark4} using MPCD simulations.

%In our case, instead, we chose two different gravity regimes, a moderate case with $\alpha=0.3$ and a strong case with $\alpha=0.06$ and explicitly showing the data also for very weak pullers and pushers for $|\beta|=0.5$. 
%Our results are consistent with this work as regards the fact that when pullers and neutral squirmers can reorient in the direction perpendicular to the plate, pushers ($\beta \leq -1$), instead, can get a tilted orientation.
%Figure \ref{no_chirality} (c) compares the stationary orientation angle of the squirmers from the present simulations for strong gravity $\alpha=0.06$ with those of Kuhr et al. in \cite{stark6}. 
%The both data coincide for $-2 \leq \beta \leq 2$ whereas differ $|\beta|=5$. 
%As previously mentioned, possible causes can be the discretization errors \cite{yamamoto3}, in solving the dynamics on the lattice, which increase at greater $\beta$ value and also the fact that our technique neglects any thermal noise which is included in their MPCD simulations.

As previously reported, the introduction of the chirality affects the trajectories of the swimming squirmers, which turn from straight, in the absence of chirality, into curved ones.
As regards the data of the curved trajectories and curvature for $\beta=-5$ at various $\chi$ in Fig. \ref{chirality_plots} (a)-(b) they are consistent with literature.
In Fig. \ref{chirality_plots} (a) the various trajectories for the strong pusher $\beta=-5$ are shown. In the case of $\chi>0$ ($\chi<0$), the pusher describes CW (CCW) trajectories. 
This is consistent with the previous work of Ishimoto et al.~\cite{ishimoto} and CW circular trajectories of \textit{E. Coli} experimentally observed near a rigid boundary \cite{e_coli,lauga2,dileonardo1,dileonardo2}. 
However, Ishimoto et al. simulated a spheroidal squirmer near a plate, taking into account the effect of chirality only for the case of a neutral squirmer $\beta=0$, and neglecting the gravity force \cite{ishimoto}. 
Instead, we performed a systematic study, including also pushers and pullers, but only considering spherical squirmers.

Finally, it was shown in Fig. \ref{chirality_plots} (b) that the linear dependency of the curvature on the chirality parameter $\kappa\propto\chi$ qualitatively agrees with the theoretical result \cite{lauga4}.

\section{\label{sec:level1_5}Summary}

In this work, we have studied the dynamics of a single sedimenting squirmer under gravity by taking into account the chirality via the rotlet dipole term in the surface velocity. Different dynamics emerge upon varying the gravity strength. In the absence of chirality, for strong gravity, all types of squirmers sediment to the bottom plate and reorient in the perpendicular direction. For moderate gravity, instead, pushers tend to tilt from the perpendicular direction and continuously swim in the tilted direction on the bottom plate. The introduction of the chirality does not alter the stationary height and the orientation of the squirmer significantly, but distorts its trajectories, with an increasing curvature at increasing chirality values.

The present study with chiral swimmers will be extended to consider the collective behaviours of sedimenting squirmers \cite{lintuvuori1,lintuvuori2,stark5,stark6}. Combinatory applications of gravity, confinement, chirality, and hydrodynamic interactions would lead to formations of rich phenomena and new dynamical behaviours to investigate.
%Another potential study could be the introduction of non-spherical elongated squirmers, already introduced in some previous works, to give a more realistic description of real swimming microorganisms analyzing how the shape of the object can affect the dynamics of the system \cite{gompper3,gompper4}.

\section*{ACKNOWLEDGMENTS}

We are grateful to Norihiro Oyama and Matthew Turner for stimulating and enlightening discussions during the preparation of this work.
This research is supported by the Japan Society for the Promotion of Science (JSPS) KAKENHI (17H01083) grant, and the JSPS bilateral joint research projects.

\providecommand{\noopsort}[1]{}\providecommand{\singleletter}[1]{#1}%
%

%\bibliography{preref}% Produces the bibliography via BibTeX.

\end{document}